\documentclass[a4paper,11pt,pdf]{article}

\usepackage{amsfonts}
\usepackage{amsmath}
\usepackage{amsthm}
\usepackage[auth-sc,affil-sl]{authblk}
\usepackage{bm}
\usepackage{multirow}
\usepackage{verbatim}
\usepackage{enumitem}

\renewcommand
\baselinestretch{1.0} 
\usepackage{url}
\usepackage{natbib}
\bibpunct{(}{)}{,}{a}{}{,}
\usepackage{graphicx}
\usepackage{lscape}
\usepackage[usenames]{color}
\usepackage{sansmath}
\usepackage{subfigure}
\usepackage{rotating}

\DeclareMathAlphabet{\mathpzc}{OT1}{pzc}{m}{it}

\newcommand\independent{\protect\mathpalette{\protect\independenT}{\perp}}
\def\independenT#1#2{\mathrel{\rlap{$#1#2$}\mkern2mu{#1#2}}}

\def\Expect{\mathbb{{E}}}
\def\P{\text{Pr}}
\newcommand \address[1]{\gdef \@address{#1}}
\makeatletter
\long\def\@footnotetext#1{\insert\footins{\def\baselinestretch{1.2}\footnotesize
\interlinepenalty\interfootnotelinepenalty
\splittopskip\footnotesep \splitmaxdepth \dp\strutbox
\floatingpenalty \@MM \hsize\columnwidth \@parboxrestore
\edef\@currentlabel{\csname
p@footnote\endcsname\@thefnmark}\@makefntext
{\rule{\z@}{\footnotesep}\ignorespaces #1\strut}}}

\long\def\symbolfootnote[#1]#2{\begingroup%
\def\thefootnote{\fnsymbol{footnote}}\footnote[#1]{#2}\endgroup}

\providecommand{\keywords}[1]{\textit{Keywords:} #1}

\def\maketitle{%
  \null
  \thispagestyle{empty}%
  \begin{center}\leavevmode
    \normalfont
    {\LARGE \bf \@title\par}%
    {\normalsize \@author\par}%
    \vskip 0.05 cm
    \vskip 0.05cm
    {\normalsize \@date\par}%
  \end{center}%
}

\makeatletter
\newcommand{\institute}[1]{\newcommand{\@institute}{#1}}

\renewcommand\textsl{\textcolor{blue}}

\begin{document}
\title{Quantifying the causal effect of speed cameras on road traffic collisions via an approximate Bayesian doubly robust estimator}
\author[1]{Daniel J. Graham}
\affil[1]{Corresponding author: Department of Civil Engineering, Imperial College London, London, SW7 2AZ, UK. Email: \texttt{d.j.graham@imperial.ac.uk}} 
%\affil[ ]{Email: \texttt{d.j.graham@imperial.ac.uk}}
%
\author[2]{Cian Naik}
\author[2]{Emma J. McCoy}
\affil[2]{Department of Mathematics, Imperial College London, London, UK}
\author[3]{Haojie Li}
\affil[3]{School of Transportation, Southeast University, Nanjing, China}

\date{}

\maketitle
\begin{abstract}
This paper quantifies the effect of speed cameras on road traffic collisions using an approximate Bayesian doubly-robust (DR) causal inference estimation method. Previous empirical work on this topic, which shows a diverse range of estimated effects, is based largely on outcome regression (OR) models using the Empirical Bayes approach or on simple before and after comparisons. Issues of causality and confounding have received little formal attention. A causal DR approach combines propensity score (PS) and OR models to give an average treatment effect (ATE) estimator that is consistent and asymptotically normal under correct specification of either of the two component models. We develop this approach within a novel approximate Bayesian framework to derive posterior predictive distributions for the ATE of speed cameras on road traffic collisions. Our results for England indicate significant reductions in the number of collisions at speed cameras sites (mean ATE = -15\%). Our proposed method offers a promising approach for evaluation of transport safety interventions.
\end{abstract}
.

\section{Introduction}

Fixed speed limit enforcement cameras are a common intervention used to encourage drivers to comply with maximum legal speed limits. The cameras are installed at sites on selected links in order to detect speed limit violations, which can subsequently be punished with monetary fines, driver licence disqualification points, or prosecution. Since the introduction of speed cameras (SCs) there has been considerable debate about their effects on road traffic collisions (RTCs). At various times claims have been made that SCs serve to reduce RTCs, that they have no effect, or even that they increase RTCs by encouraging more erratic driving behaviour.  

A number of academic studies of the effect of speed cameras on RTCs have been undertaken \citep[for a review see][]{Li/et/al:2013}. Most studies find that speed cameras have led to a reduction in RTCs, but the range of estimated effects is large (from 0\% to -55\%). Variation in estimates is to be expected given that study results pertain to diverse empirical contexts, but it is also the case that a number of different methods have been applied which can have a critical influence on results obtained. In particular, since SCs are not randomly assigned, it is essential that any adopted method recognises that the observed relationship between SCs on RTCs may be subject to confounding. Confounding arises when the characteristics that influence treatment assignment (i.e. whether a site is `treated' and `untreated' with an SC) also matter for outcomes (i.e. RTCs). Regression to the mean (RTM), for instance, is a well known manifestation of confounding that arises via `selection bias'. 

The extent to which confounding has been recognised and addressed in existing studies varies considerably. Some studies have simply ignored it, using simple before-and-after methods with control groups \citep[e.g.][]{Christie/et/al:2003, Cunningham/et/al:2008, DePauw/et/al:2014, Gains/et/al:2004, Gains/et/al:2005, Goldenbeld/vanSchagen:2005, Jones/et/al:2008, Maher:2015}. Others have used the empirical Bayes (EB) method as suggested by \citet{Hauer/et/al:2002}, largely to adjust for effects of confounding that arise via RTM \citep[e.g.][]{Chen/et/al:2002, Elvik:1997, Hoye:2015, Mountain/et/al:2004, Mountain/et/al:2005, Shin/et/al:2009}. Finally, there are a small number of studies that have used time-series methods, either interrupted time-series analyses with control groups or ARIMA, to test for changes in outcome rates \citep[e.g.]{Carnis/Blais:2013, Hess/Polak:2003, Keall/et/al:2001}. Where studies have attempted to address confounding this has been done via the inclusion of covariates in outcome regression (OR) models, typically using Poisson or negative binomial Generalised Linear Models (GLMs). 

In a previous paper we adopted a propensity score (PS) matching approach to evaluate the effectiveness of speed cameras \citep[see][]{Li/et/al:2013}. A key advantage of the PS over OR approach is that it provides an effective way of isolating a valid control group by ensuring that the distribution of pre-treatment covariates matches those of the treated group and that genuine overlap in the support of the covariates exists between the two groups. However, as with the OR approach, valid inference from PS models crucially depends on the unknown PS model being correctly specified.

In this paper we build on our previous work by developing and applying an estimation approach which we believe has much to offer in evaluating the effectiveness of road safety interventions. Our approach uses the principle of doubly-robust (DR) estimation, which provides robustness to model misspecification by combining both OR and PS models to derive an average treatment effect (ATE) estimator which is consistent and asymptotically normal under correct specification of just one of the two component models. The DR approach is attractive for our application because the PS and OR models we can construct make different assumptions about the nature of confounding. For the PS model, we are able to faithfully represent via measured covariates the formal criteria that exist for the assignment of speeds cameras to sites. For the OR model, we can difference our response variable before and after treatment to allow for the existence of site level time-invariant unobserved effects in addition to measured confounders. 

To avoid common sources of misspecification error, we estimate our component models using semiparametric Generalized Additive Mixed Models (GAMMs) which make minimal a priori assumptions on the functional form of the relationships under study. We also use a matching algorithm prior to forming the DR model to establish a valid control group. Thus, in our approach, potential biases from confounding are addressed by combining three compatible modelling tools: via matching to achieve comparability between treated and control sites, via a regression model for RTCs, and via a model for the treatment assignment mechanism.    

DR estimators have been studied and applied extensively in the frequentist setting \citep[e.g.][]{Robins:2000,Robins/et/al:2000c,Robins/Rotnitzky:2001,VanDerLaan/Robins:2003,Lunceford/Davidian:2004,Bang/Robins:2005,Kang/Schafer:2007}. A further contribution of the paper is that we develop our binary DR estimator within the Bayesian paradigm. A Bayesian representation of the DR model has proven difficult to formulate in previous work because DR estimators are typically constructed as solutions to estimating equations based on a set of moment restrictions that do not imply fully specified likelihood functions. We choose the Bayesian paradigm for three main reasons. First, DR estimation of the ATE involves prediction and extrapolation over covariate distributions with underlying uncertainty in parameter estimates. Bayesian inference provides a suitable framework for prediction that explicitly addresses such uncertainty in the sense that both the predicted observations, and the relevant parameters for prediction, have the same random status. Second, by deriving a posterior predictive distribution for the ATE, rather than a fixed value, we can make probability statements about the causal quantity of interest allowing us to discuss findings in relation to specific hypotheses or in terms of credible intervals which can offer a more intuitive understanding of the effects of SCs for public policy formulation. Finally, we develop an approximate Bayesian approach that can utilise prior information about the parameters of interest, which could be useful in evaluating safety interventions when historical data or training data from other regions are available. 

The paper is structured as follows. Section two outlines broad trends in road traffic casualties for Britain and then sets out a formal causal modelling framework to estimate the effects of SCs on RTCs. Section three describes our approximate Bayesian DR approach and presents some simulations that demonstrate its properties. Section four describes the data available for estimation and outlines our chosen model specifications. Results are then presented in section five and conclusions are drawn in the final section.

\section{A causal inference framework to quantify the effects of speed cameras}

\subsection{Road traffic casualties in Britain}

For the year ending September 2016 the UK DfT recorded a total of 182,560 causalities on British roads of which 25,160 were classified as killed or seriously injured (KSIs) \citep[][]{DfT:2017}. Since 2010 the annual numbers of fatalities and KSIs have not changed significantly, following several years in which road safety was improving. The average number of fatal road traffic incidents over the period 2010 to 2016 is approximately 1,800. Since the volume of road traffic has continued to grow over this period, however, the number of fatalities per vehicle mile driven has been falling \citep[][]{DfT:2016}.

The DfT argue that there is good evidence to suggest that while the absolute number of fatalities on British roads now appears to be relatively static, overall absolute casualty numbers are continuing to fall. In  short, levels of safety appear to be improving in relative terms and not deteriorating in absolute terms. Given the changes that have occurred in vehicle technology, medical care, and road safety interventions, however, the DfT also note that a comprehensive causal understanding of the factors underpinning casualty trends is currently out of reach. In this paper we attempt to contribute to such an understanding by quantifying the causal impact of one type of safety intervention: speed cameras (SCs).               

\subsection{ATE estimation within the potential outcomes framework}

Our sample comprises $n$, $i=1,...,n$, links on the road network. Some links have a SC other do not. We define $D_i\in\{1,0\}$ as a binary random variable indicating the presence or otherwise of a SC and we refer to this as the treatment variable. We are interested in the effect of the treatment on an outcome $Y_i$, which measures collision frequency. We define $Y_i(1)$ and $Y_i(0)$ as the {\em potential outcomes} for unit $i$ under treated and control status respectively. Recognising that SCs are not assigned randomly, we also define $X_i$ as a random vector of pre-treatment covariates that capture characteristics of links that are relevant to whether a SC was assigned or not, and are also relevant for outcomes. Thus, the data we observe for each link takes the form of a random vector, $z_i = (y_i,d_i,x_i)$, where $y_i$ denotes a response, $d_i$ the treatment received, and $x_i$ a vector of pre-treatment covariates.  

Ideally, we would assess the effects of SCs on each link by calculating the individual causal effect (ICE): $\tau_i=Y_i(1)-Y_i(0)$, but the observed data reveal only actual outcomes not potential outcomes. Thus we observe the random variable $Y_i=Y_i(1) I_{1}(D_i)+Y_i(0)(1-I_{1}(D_i))$, where $I_{1}(D_i)$ is the indicator function for receiving the treatment, but we do not observe the joint density, $f(Y_i(0),Y_i(1))$, since a SC cannot be both present and absent on a link simultaneously. Instead, our target of inference is the ATE, defined as
\[\tau=\Expect[Y_i(1)]-\Expect[Y_i(0)],\]
which measures the difference in expected outcomes under treatment and control status.          

A key insight of the potential outcomes approach is that if we focus on estimating the ATE then we do not have to observe all potential outcomes, even under a non-random treatment assignment, as long as three key assumptions hold. First, the potential outcomes for unit $i$ must be conditionally independent of the treatment assignment given a (sufficient) set of observed covariates $X_i$: $Y_i(0),Y_i(1)) \independent I_{1}(D_i)|X_i$. Second, the support of the conditional distribution of $X_i$ given a particular treatment status must overlap with that of $X_i$ given any other treatment status: $0<\P(I_{1}(D_i)=1|X_i=x)<1, \ \forall \ x$. Third, the relationship between observed and potential outcomes must comply with the Stable Unit Treatment Value Assumption (SUTVA) \citep[e.g.][]{Rubin:1978,Rubin:1980,Rubin:1986,Rubin:1990}, which requires that the observed response under a given treatment allocation is equivalent to the potential response under that treatment allocation: $Y_i=I_{1}(D_i)Y_i(1)+ (1-I_{1}(D_i))Y_i(0)$ for all $i=1,...,n$. 

The three assumptions defined above, which are together referred to by \citet{Rosenbaum/Rubin:1983b} as {\it strong ignorability}, allow for identification of causal effects from observational data because if they hold the ATE can be derived as,  
\begin{subequations}
\label{iden}
\begin{align}
\tau =& \Expect_i(Y_i(1)-Y_i(0))= \Expect_X\left[\Expect_i(Y_i(1)|X_i=x)-\Expect_i(Y_i(0)|X_i=x)\right]\\
=&\Expect_X\left[\Expect_i(Y_i(1)|X_i=x, I_{1}(D_i)=1)-\Expect_i(Y_i(0)|X_i=x, I_{1}(D_i)=0)\right]\\
=& \Expect_X\left[\Expect_i(Y_i|X_i=x, I_{1}(D_i)=1)-\Expect_i(Y_i|X_i=x,I_{1}(D_i)=0)\right].
\end{align}
\end{subequations}
Conditional independence justifies the equality of (\ref{iden}a) and (\ref{iden}b), the SUTVA allows the substitution of observed for potential outcomes to give (\ref{iden}c), and overlap ensures that the population ATE in (\ref{iden}c) is estimable since there are units in both the treated and untreated groups.

Thus, if strong ignorability holds, the potential outcomes approach offers a route to obtaining valid causal estimates of the ATE of SCs. To proceed we need to estimate the relevant expectations in (\ref{iden}c) above.     

\subsection{Causal estimators}
Using the notation of \citet{Tsiatis/Davidian:2007}, we define joint densities of the observed data of the form
\[f_Z(z)=f_{Y|D,X}(y|d,x)f_{D|X}(d|x)f_X(x).\] 
Given strong ignorability, estimation of the ATE of SCs can proceed in one of the following ways;
\begin{enumerate}
\item[i.] Outcome regression (OR) model - leave $f_{D|X}(d|x)$ and $f_{X}(x)$ unspecified and posit a model for $\Expect[Y_i|D_i,X_i])$; the mean of the conditional density of the response given the covariates, using an OR model $\Psi^{-1}\{m(D_i,X_i;\beta)\}$, for known link function $\Psi$, regression function $m()$, and unknown parameter vector $\beta$. If the OR is correctly specified for the mean response then the ATE can be consistently estimated by.
\[\hat{\tau}_{OR}=\frac{1}{n}\sum_{i=1}^n\left[\Psi^{-1}\{m(1,X_i; \hat{\beta})\}-\Psi^{-1}\{m(0,X_i; \hat{\beta})\}\right].\]   
\item[ii.] Propensity score (PS) model - leave $f_{Y|D,X}(y|d,x)$ and $f_X(x)$ unspecified but assume a model for $f_{D|X}(d|x)$; the conditional density of treatment assignment given covariates. This is a propensity score (PS) model, denoted $\pi(D_i|X_i; \alpha)$, which can be used to form a number of different nonparametric estimators but of primary interest here is its use in the weighting estimator attributed to \citet{Horovitz/Thompson:1952}
\[\hat{\tau}_{IPW}=\frac{1}{n}\sum_{i=1}^n \left[\frac{I_{1}(D_i) \cdot Y_i}{\pi(D_i|X_i; \hat{\alpha})} - \frac{[1-I_{1}(D_i)] \cdot Y_i}{1-\pi(D_i|X_i; \hat{\alpha})}\right],\] 
which is consistent under correct specification of the PS by virtue of the fact that $\Expect[Y_i(1)]=\Expect\{\left[Y_i(1) \cdot I_{1}(D_i)\right]/\pi(D_i|X_i; \alpha)\}$ and similar for control treatment status. 
\item[iii.] Doubly-robust (DR) model - leave $f(x)$ unspecified but assume both an OR model and a PS model and combine them to form a DR estimator. This is achieved by weighting or augmenting the OR model with a function of the inverse of the estimated PS to give a DR model. In this paper we estimate the weighted model
\[e(D_i,X_i;\xi)=\Psi^{-1}\{m(D_i,X_i; \xi)\}\]
where the unknown parameter vector $\xi$ is obtained by weighting the model with
\[\widehat{\kappa}_i(D_i,X_i)=\frac{I_{1}(D_i)}{\widehat{\pi}(D_i| X_i; \widehat{\alpha})}+\frac{1-I_{1}(D_i)}{1-\widehat{\pi}(D_i| X_i; \widehat{\alpha})}.\] 
This model will consistently estimate $\Expect[Y_i|D_i,X_i])$ if the model $\Psi^{-1}\{m(I_{1}(D_i),X_i;\beta)\}$ is correct because while weighting may induce inefficiency it will leave the consistency and asymptotic normality of the OR estimates unchanged. If the OR model is incorrectly specified, but the PS is correctly specified, the model is still consistent because weighting gives rise to estimating equations of the form
\begin{equation}
\sum_{i=1}^{n}\widehat{\kappa}_i(D_i,X_i)\frac{1}{\phi} \frac{\partial e\left(d_i,x_i;\xi\right)}{\partial \xi^{\sf{T}}} \left[y_i-e\left(d_i,x_i;\xi\right)\right]=0,
\end{equation}
where $\phi_i \equiv \phi (D_i,X_i)$ is a working conditional variance for $Y_i$ given $(D_i,X_i)$, which effectively correct for the bias in approximating $\Expect[Y_i|D_i,X_i]]$ using $\Psi^{-1}\{m(D_i,X_i;\beta)\}$ \citep[for a proof see][] {Lunceford/Davidian:2004}.   

We use estimates of $\xi$ to form the DR estimator 
\[\hat{\tau}_{DR}=\frac{1}{n}\sum_{i=1}^n\left[\Psi^{-1}\{m(1,X_i; \hat{\xi})\}-\Psi^{-1}\{m(0,X_i; \hat{\xi})\}\right].\]
\end{enumerate}

\section{Approximate Bayesian doubly-robust estimation}

So far we have discussed DR estimation within the context of frequentist semiparametric inference. As mentioned in the introduction to the paper there are good reasons why a Bayesian inferential approach is particularly beneficial for estimation of road safety interventions. Bayesian inference has, however, proven difficult to apply for DR estimators because they are based on a set of moment restrictions which do not provide fully specified likelihood functions. Here, we make some improvements to the approach proposed by \citet{Graham/et/al:2016} in the context of continuous treatment. In contrast to that paper we focus on binary treatments using PS weighting rather than augmentation to achieve the DR model and we implement ways of incorporating prior information into the posterior distribution of the ATE. The basic theory underpinning approximate Bayesian inference in this context is covered comprehensively in \citet{Graham/et/al:2016} and so we provide only a brief summary here.

The Bayesian bootstrap was first introduced by \citet{Rubin:1981} and applied in weighted likelihood models by \citet{Newton/Raftery:1994}. The basic idea is to create new datasets by repeatedly re-weighting the original data in order to obtain the posterior distribution for some parameter of interest. If we treat our observed data, $z_i$ say, as effectively coming from a multinomial distribution with distinct values $a_k$, $k=(1,...,K)$, and attach a probability to each distinct value $\theta=(\theta_1,...,\theta_k)$, then by placing an improper Dirichlet prior on $\theta$
\[\pi(\theta)\propto \prod^{K}_{k=1}\theta_{k}^{-1}.\]
the posterior density also has a Dirichlet distribution
\[p(\theta|v)\propto \prod^{K}_{k=1}\theta_{k}^{n_{k}-1}.\]
with parameter $n_{k}$. This posterior can be stimulated via the weighted likelihood
\[\widetilde{L}(\theta)=\prod^{n}_{i=1}f(z_i;\theta)^{w_{i}^{ }},\]
in which the weights $w^{ }= (w_{1}^{ },...,w_{n}^{ } )$ are distributed according to the uniform Dirichlet distribution and simulated as $n$ independent standard exponential (i.e. gamma(1,1)) variates and standardised.  The weighted likelihood reduces to
\[\widetilde{L}(\theta)= \prod_{i=1}^n \left\{ \prod_{k=1}^K \theta_k^{I_{k}(z_i)} \right\}^{w_i} = \prod_{k=1}^{K}\theta_{k}^{\sum\limits_{i=1}^n w_i I_{k}(z_i)} =  \prod^{K}_{k=1}\theta_{k}^{n \gamma_{k}},\]
say, where $n \gamma_{k}$ is the sum of the weights $w_i$ for which $z_i = a_k$. Since the vector $\gamma=(\gamma_1,...,\gamma_K)$ has a Dirichlet distribution with parameters $n_k=(n_1,...,n_K)$,
\[p(\gamma)\propto \prod_{k=1}^K\gamma_k^{n_k-1}\]
and since at the point of maximisation of $\widetilde L (\theta)$ is $\widetilde \theta = \gamma$, then the solutions to the maximised weighted likelihood function with repeatedly sampled uniform Dirichlet weights $w^{(l)}$ represent a sample from the posterior of $\theta$ under the improper prior $\prod_k \theta^{-1}_{k}$.

To apply the Bayesian bootstrap to our DR model we estimate
\[e(D_i,X_i;\xi)=\Psi^{-1}\{m_A(D_i,X_i; \xi)\}\]
with weights
\[w_i^{(l)} \cdot \widehat{\kappa}_i(D_i,X_i).\]
The maximiser of $\widetilde{L}(\xi)$, which we denote $\widetilde{\xi}$, implies a solution to
\begin{equation}\label{BBAPO}
\sum_{i=1}^{n}w^{(l)}_{i}\cdot \widehat{\kappa}_i(D_i,X_i) \cdot \frac{1}{\phi} \frac{\partial e\left(d_i,x_i;\xi\right)}{\partial \xi^{\sf{T}}} \left[y_i-e\left(d_i,x_i);\xi\right)\right]=0,
\end{equation}
which as noted above has the DR property. We repeatedly draw sets of random weights $\{w^{(l)}_i\}^{n}_{i=1}$ as $n$ standardised independent standard exponential variates and solve (\ref{BBAPO}) to build up an empirical posterior density of $\widetilde \xi$, denoted $p_n(\widetilde \xi)$, from which the sampled values $\widetilde \xi^{(l)}$ are consistent with the DR estimating equations.

\citet{Newton/Raftery:1994} apply sampling-importance resampling (SIR) to improve accuracy of the weighted bootstrap approach, but this improvement requires a fully specified likelihood function. Instead, for our restricted moment model, we use the resampling scheme proposed by \citet{Muliere/Secchi:1996} which extends Rubin's bootstrap in a general Bayesian nonparametric context. Two attractive features of Muliere and Secchi's approach for causal modelling are that it ensures that predictive distributions are not constrained to be concentrated on observed values and it allows us to take prior opinions into account. The posterior predictive distribution of the ATE, incorporating prior information, is obtained in the following way.  

\begin{enumerate}
\item[i.] Estimate the PS model $\pi(D_i|X_i; \alpha)$, and form 
\[\widehat{\kappa}_i\left(d_i,x_i;\widehat{\alpha}\right)=\frac{I_1(d_i)}{\widehat{\pi}\left(d_i| x_i; \widehat{\alpha}\right) }+\frac{1-I_1(d_i)}{1-\widehat{\pi}\left(d_i| x_i; \widehat{\alpha}\right) }.\] 
\item[ii.] Draw a single set of random weights $\{w^{(l)}_i\}^{n}_{i=1}$ and form the combined weights $w_i^{(l)} \cdot \widehat{\kappa}_i\left(d_i| x_i; \widehat{\alpha}\right)$ and estimate the weighted model 
\[
\Psi^{-1}\left\{m_A\left(d_i,x_i; \xi^{(l)} \right)\right\}.
\]
\item[iii.] Repeatedly compute (ii) using new weights $\{w^{(l)}_i\}^{n}_{i=1}$ to obtain the empirical posterior distribution $p_n(\widetilde \xi)$.
\item[iv.] Introduce a prior distribution $p_0$ for $\xi$ and a positive number $k$, the `measure of faith' that we have in this prior. This can range anywhere from 1 to a size comparable to the number of samples of $\xi$. 
\item[v.] Generate $m$ observations $x_1^*,...,x_m^*$ from $\frac{kp_0+Lp_n}{k+L}$, where $p_n$ is as above. We choose $m=L$ in our case.
\item[vi.] For $i=1,...,m$ generate $v_i$ from a $\Gamma\bigg(\frac{L+k}{m},1\bigg)$ distribution.
\item[vii.] Sample new parameters $\widetilde \xi_{MS}$ from $x_1^*,...,x_m^*$ using the weights $v_1,...,v_m$ to form the posterior $p_m(\widetilde \xi)$.
\item[viii.] Resample $V$ values of the covariate vector uniformly over the observed values and a single vector $\xi^{(m)}$ from $p_m(\widetilde \xi)$.
\item[ix.] Form a sampled value of the ATE random variable as
\[
\tau^{(m)}_{BDR}= \frac{1}{V}\sum_{v=1}^{V} \left[\Psi^{-1} \{m_A(1, x_v; \widetilde{\xi}^{(m)})\}-\Psi^{-1} \{m_A(0, x_v; \widetilde{\xi}^{(m)})\}\right] .
\]
\item[x.] Repeat this procedure $M$ times, $m=(1,...,M)$, to obtain the posterior predictive distribution.
\end{enumerate}

\subsection{Simulations}\label{C6S4}
In this subsection we present some simulation to demonstrate the DR properties of our approximate Bayesian approach. The simulations are based on the following data generating process: a binary treatment $D$ is assigned as a function of covariate $X$, and the outcome of interest $Y$ depends on both treatment $D$ and covariate $X$
\begin{eqnarray*}
X \sim \text{Normal}(0,10)\\
D \sim \text{Bernoulli}(\text{expit}(\alpha_0+\alpha_1 X))\\
Y \sim \text{Normal}(\beta_0 + \beta_1 D + \beta_2 X, 5)
\end{eqnarray*}
where $\alpha_0=2$, $\alpha_1=0.2$, $\beta_0=10$, $\beta_1=5$, $\beta_2=0.2$. The true ATE is given by parameter $\beta_1$, that is $\tau = 5.0$.

The following models are tested:
\begin{itemize}
\item[1.] $\widehat{\tau}_{BOR1}$ - an approximate Bayesian OR model based on the correctly specified model: $\Expect[Y|D,X]=\beta_0+\beta_{1}D+\beta_{2}X$. The point estimate reported in the simulations is the mean value of the ATE posterior predictive distribution, i.e.   
\[
\widehat{\tau}_{BOR}= \frac{1}{L}\sum_{l=1}^{L}\left[\frac{1}{V}\sum_{v=1}^{V} \left[\Psi^{-1} \left\{m\left(1, x_v; \widetilde{\beta}^{(l)}\right)\right\}-\Psi^{-1} \left\{m\left(0, x_v; \widetilde{\beta}^{(l)}\right)\right\}\right] \right].
\]
\item[2.] $\widehat{\tau}_{BOR2}$ - same as [1.] except based based on an incorrectly specified OR model with covariate $X$ excluded.
\item[3.] $\widehat{\tau}_{PS1}$ - an approximate Bayesian inverse PS weighted model based on the correctly specified PS model
\[
\widehat{\tau}_{PS1}= \frac{1}{L}\sum_{l=1}^{L}\left[\frac{1}{V}\sum_{v=1}^{V} \left[y_v \cdot \frac{d_v-\widehat{\pi}(d_v|x_v; \widetilde{\alpha}^{(l)})}{\widehat{\pi}(d_v|x_v; \widetilde{\alpha}^{(l)})(1-\widehat{\pi}(d_v|x_v; \widetilde{\alpha}^{(l)}))}\right] \right]
\]
\item[4.] $\widehat{\tau}_{PS2}$ - an approximate Bayesian inverse PS weighted model based on an incorrectly specified PS model, in which the PS is generated randomly from the continuous uniform distribution: $\widehat{\pi}(D|X) \sim \text{Uniform}(0,1)$.
\item[5.] $\widehat{\tau}_{BDR1}$ - an approximate Bayesian DR model based on an incorrectly specified OR model ($X$ excluded) but with weights based on the correct PS model
\[
\widehat{\tau}_{BDR}= \frac{1}{L}\sum_{l=1}^{L}\left[\frac{1}{V}\sum_{v=1}^{V} \left[\Psi^{-1} \left\{m_A\left(1, x_v; \widetilde{\xi}^{(l)}\right)\right\}-\Psi^{-1} \left\{m_A\left(0, x_v; \widetilde{\xi}^{(l)}\right)\right\}\right] \right].
\]
\item[6.] $\widehat{\tau}_{BDR2}$ - an approximate Bayesian DR model based on a correctly specified OR model but with weights based on the incorrect PS model.
\item[7.] $\widehat{\tau}_{BDR3}$ - an approximate Bayesian DR model based on the incorrectly specified OR model weighted with weights based on the incorrect PS model.
\end{itemize}

The simulations are based on 1000 runs on generated datasets of size 1,000. In each case, we place a Normal prior on the treatment coefficient $\beta_1$, with mean equal to the true value (5 in this case). We set the measure of faith $k$ to be relatively low so as not to overly affect the results.  Table \ref{sims} shows our simulation results.  Mean values and variances of the point estimates obtained (i.e. means and variances of the ATE distributions) and the mean squared error (MSE) are reported.

\begin{table}[htbp]
\centering
{\bf \caption{Simulation results for posterior predictive distributions ($\tau=5.0$). \label{sims}}}
\begin{tabular}{lccc}
\hline
	& Av. Est. & Emp. Var. & MSE \\
\hline
BOR1	& 5.004 & 0.036 & 0.036 \\
BOR2	& 5.350 & 0.036 & 0.157 \\
PS1	& 4.998 & 0.118 & 0.119 \\
PS2	& 276.740 & 1.84E+07 & 1.85E+07 \\
BDR1	& 5.008 & 0.046 & 0.046 \\
BDR2	& 5.018 & 0.862 & 0.862 \\
BDR3	& 5.360 & 0.946 & 1.074 \\
\hline
\end{tabular}
\end{table}

The mean of the posterior distribution for the ATE from the correctly specified OR model, $\widehat{\tau}_{BOR1}$, provides a good approximation to the true value of $\tau$. The incorrectly specified OR model, BOR2, fails to address confounding and consequently $\widehat{\tau}_{BOR2}$ provides a poor approximation to the true ATE. A good estimate of $\tau$ is achieved via the correctly specified PS model ($\widehat{\tau}_{PS1}$), but when the PS is model is mispecified ($\widehat{\tau}_{PS1}$) the estimate of the ATE is far away from the true value. In our simulations the PS model is severely misspecified, or simply wrong, having being generated randomly. This tendency of the inverse PS model to fail quite considerably under severe misspecification is well known in the literature \cite{Kang/Schafer:2007}.  Weighting the incorrectly specified OR model with weights $\widehat{\kappa}(D,X)$, based on a correctly specified PS model, as in the BDR1 model, provides correction for misspecification bias with an average point estimate very close to the true value, but slightly larger posterior variances relative to the correctly specified OR model. The BDR2 model simulation also produces valid point estimates because weighting by weights based on an incorrectly specified PS model does not does not induce bias, but it does increase variance. Finally, if both the OR and PS models are wrongly specified as in BDR3, the model fails to produce a good point estimate of the mean ATE.  

\section{Data and model specifications}   

\subsection{Treatment and outcome variable}

We have data on the location of fixed speed cameras for 771 camera sites in the following English administrative districts: Cheshire, Dorset, Greater Manchester, Lancashire, Leicester, Merseyside, Sussex and the West Midlands. These sites form our group of treated units. To select potential control sites we randomly sampled a total of 4787 points on the network within our eight administrative districts. The large ratio of potential control to treated units is adopted to ensure that we have a sufficient number of control units after we apply a matching algorithm.    

Our outcome variable is the number of personal injury collisions (PICs) per kilometre as recorded from the location of the speed cameras, or in the case of control groups, from the randomly selected point. The PIC data are taken from records completed by police officers each time that an incident is reported to them. The individual police records are collated and processed by the UK Department for Transport as the `STATS 19' data. The location of each PIC is recorded using the British National Grid coordinate system and can be located on a map using Geographical Information System (GIS) software. Because the established dates of speed cameras vary from 2002 to 2004, the period of analysis is from 1999 to 2007 to ensure the availability of collision data for the years before and after the camera installation for every camera site.

\subsection{Covariates}

To adequately adjust for confounding we require a set of measured covariates that adequately represent the characteristics of units that simultaneously determine treatment assignment and outcome. For the UK there exists a formal set of site selection guidelines for fixed speed cameras \citep[see][]{Gains/et/al:2004} that are extremely valuable in choosing covariates. The criteria are as follows
\begin{enumerate}
\item Site length: between 400-1500 m.
\item Number of fatal and serious collisions (FSCs): at least 4 FSCs per km in last three calendar years.
\item Number of personal injury collisions (PICs): at least 8 PICs per km in last three calendar years.
\item 85th percentile speed at collision hot spots: 85th percentile speed at least 10\% above speed limit.
\item Percentage over the speed limit: at least 20\% of drivers are exceeding the speed limit.
\end{enumerate}

Criteria one to three are primary guidelines for site section and criteria four and five are of secondary importance. There are sites that do not meet the above the above criteria that will still be selected as enforcement sites, mainly for reasons such as community concern and engineering factors.

Selection of the speed camera sites was primarily based on collision history. collision data can be obtained from the STATS 19 database and located on the map using GIS. However, secondary criteria such as the 85th percentile speed and percentages of vehicles over the speed limit are normally unavailable for all sites on UK roads. If speed distributions differ between the treated and untreated groups, then the failure to include the speed data could bias the estimation, an issue discussed in previous research \citep[e.g.][]{Mountain/et/al:2005,Gains/et/al:2004}. For untreated sites with the speed limit of 30 mph and 40 mph, the national average mean speed and percentages of speeding are similar to the data for the camera sites. The focus groups for this study are sites with the speed limit of 30 mph and 40 mph throughout the UK. It is reasonable to assume that there is no significant difference in the speed distribution between the treated and untreated groups and hence exclusion of the speed data will not affect the accuracy of the propensity score model.

It is also possible that drivers may choose alternative routes to avoid speed cameras sites. collision reduction at camera sites may include the effect induced by a reduced traffic flow. The benefits of speed cameras will therefore be overestimated without controlling for the change in traffic flow. The annual average daily flow (AADF) is available for both treated and untreated roads and the effect due to traffic flow is controlled for in this study by including the AADF in the propensity score model. 

In addition to the criteria that strongly influence the treatment assignment, factors that affect the outcomes should also be taken into account when the propensity score model is specified. We further include road characteristics such as: road types, speed limit, and the number of minor junctions within site length, which are suggested as important factors when estimating the safety impact of speed cameras \citep[][]{Gains/et/al:2005,Christie/et/al:2003}.

\subsection{Component model specifications}

The outcome variable of interest is the number of collisions per site. For the OR model the response is specified in differenced form, i.e. the number of collisions in the post-treatment period minus the number of collisions in the pre-treatment period. Differencing allows for the existence of unit level time-invariant effects, which could be random or fixed. The PS model is estimated using a logit Generalized Additive Mixed Model (GAMM) specification. Matching and overlap is achieved using nearest neighbour matching via the {\tt MatchIt} package in {\tt R}. The weighted OR model is then estimated on the trimmed dataset, which satisfies matching and overlap conditions, using a Gaussian GAMM specification. We use GAMMs to avoid making a-priori assumptions on the functional form of the relationships under study.   

As mentioned in the introduction, the DR approach is particularly attractive for our application because of the differences inherent in our PS and OR model specifications. Due to the existence of formal criteria for SC assignment we have a high degree of confidence in the ability of our covariates to eliminate confounding via the PS model. For the OR model, differencing of the response variable before and after treatment allows for the existence of site level time-invariant unobserved effects in addition to measured confounders. Thus, there are subtle differences in the way we model the ATE via the PS or OR approaches. A degree of robustness is offered using a DR approach since we will obtain a consistent estimate of the ATE if just one of the component models is well specified.        

\section{Results}
The objective of our application is to estimate the {\em marginal} effect of SCs on RTCs, having adjusted for baseline confounders. We estimate the following models: an OR model, an IPW model, a DR model comprising an OR model weighted with the inverse PS covariate (DR), and a na{\"i}ve model which is simply the OR model without covariates. For the na{\"i}ve model we report results using the matched and full samples. All models are repeatedly estimated using the approximate Bayesian approach outlined above. In addition to the posterior predictive distribution for the ATE we report point estimates at the mean of the posterior. For comparison, we also report Frequentist results. 

The results are shown in table 2 below including means and credible intervals of the ATE distributions. Our causal models (OR, IPW and DR) indicate that the presence of speed cameras corresponds with an average change in the number of RTCs of -14.4\% to -15.5\% . Note that the approximate Bayesian and Frequentist point estimates are very similar, which is what we would expect for linear models with uninformative priors. In comparison, the Na{\"i}ve model which does not adjust for confounding, finds a higher ATE of -17.6\% using the matched sample and -33.6\% using the unmatched sample. Figure \ref{fig1} below shows the posterior predictive distribution derived from the DR model. 

\begin{table}[htbp]
\centering
{\bf  \caption{Bayesian and Frequentist bootstrapped estimates of the average treatment effect}}
\label{results}
\begin{tabular}{lccccc}
    \hline
          & \multicolumn{3}{c}{Bayesian bootstrap} & \multicolumn{2}{c}{Frequentist bootstrap} \\
    \cline{2-6}
          & posterior mean & s.d.  & 95\% cred. int.   & Est.  & s.e. \\
    \hline
OR                 & -14.429 & 3.536 & (-21.473, -7.350) &-14.825& 3.419\\
IPW                & -15.476 & 5.480 & (-26.367, -4.808) &-15.504& 4.935\\
DR                & -14.359 & 3.605 & (-21.841, -7.352) &-14.370& 3.494\\
Na{\"i}ve (matched sample)         & -17.617 & 5.199 & (-28.214, -7.800) &-17.887& 5.118\\
Na{\"i}ve (full sample)         & -33.684 & 5.911 & (-42.133, -13.624) &-34.682& 3.573\\
\hline
\end{tabular}
\end{table}
  
\begin{figure}
\centering
{\includegraphics[scale=0.4]{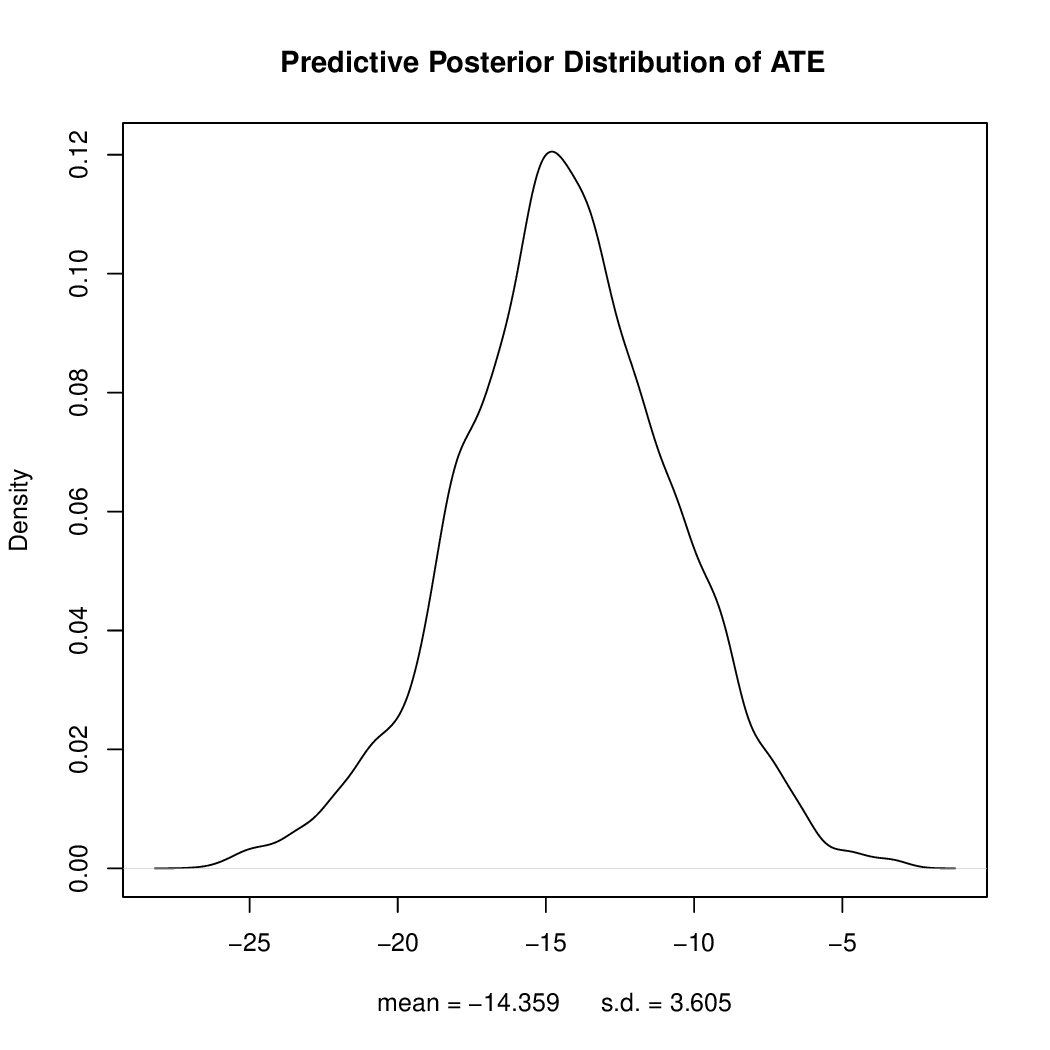} }
   {\bf \caption{Predictive posterior distribution of the average treament effect from the doubly-robust model. \label{appfig}}}
\end{figure}

Thus, it would appear that correcting for potential sources of confounding serves to reduce the magnitude of our ATE estimates, but we still find a substantial reduction in RTCs associated with presence of speed cameras. The difference in estimated ATE between the na{\"i}ve and causal models makes sense given that the formal criteria used to assign SCs favours sites that have exhibited high rates of collisions in the past. Crucially, our causal models imply that SCs do make a real difference to RTCs over and above the modelled effect of confounding from non random assignment.  

\section{Conclusions}
In this paper we have the quantified the causal effect of speed cameras on road traffic collisions via an approximate Bayesian doubly robust approach. This is the first time such an approach has been applied to study road safety outcomes. The method we propose could be used more generally for estimation of crash modification factor (CMF) distributions. Simulations demonstrate that the approach is doubly-robust for average treatment effect estimation. Our results indicate that speed cameras do cause a significant reduction in road traffic collisions, by as much as 15\% on average for treated sites. This is an important result that could help inform public policy debates on appropriate measures to reduce RTCs. The adoption of evidence based approaches by public authorities, based on clear principles of causal inference, could vastly improve their ability to evaluate different courses of action and better understand the consequences of intervention. 

There are thus two important implications of our study that could ultimately improve highway safety. First, is that such inference could be employed to achieve a more effective assignment of SCs and consequent reduction of RTCs. Second, the approach outlined above could be used to continually monitor SC effectiveness as baseline conditions (e.g. related to road traffic and wider demographic and social characteristics) change, thus providing a mean of monitoring the effectiveness of road safety interventions dynamically.          

\bibliographystyle{chicago}
\bibliography{SC-ABDR}

\end{document}